# Competing ferromagnetic and antiferromagnetic interactions drive the magnetocaloric tunability in $Gd_{55}Co_{30}Ni_xAl_{15-x}$ microwires

Yunfei Wang[1,2], Nguyen Thi My Duc[2,3], Tangfeng Feng[1], Huijie Wei[1],
Faxiang Qin,[1,*] and Manh-Huong Phan[2,*]

[1]Institute for Composites Science Innovation (InCSI), School of Materials Science and Engineering, Zhejiang University, 38 Zheda Road, Hangzhou, 310027, PR. China

[2]Department of Physics, University of South Florida, Tampa, Florida 33620, USA

[3]The University of Danang, University of Science and Education, 459 Ton Duc Thang, Lien Chieu, Danang, Vietnam

We have employed $Gd_{55}Co_{30}Ni_xAl_{15-x}$ ($x$ = 10, 5 and 0) amorphous microwires as a model system to unravel the impact of multiple magnetic interactions on the magnetism and the magnetocaloric behavior in Gd-alloy microwire systems. Our study shows that in addition to the RKKY ferromagnetic (FM) interaction (Gd-Gd), antiferromagnetic (AFM) interactions (Gd-Co, Gd-Ni) coexist and contribute to the magnetic and magnetocaloric response of the system. The dilution effect of Al element on the FM Gd-Gd interaction is responsible for the decrease of the Curie temperature ($T_C$), whereas the increase of the saturation magnetization ($M_S$) is originated from the reduced AFM Gd-Ni interaction. A thorough analysis of critical exponents suggests that the presence of the AFM interactions hinders the system to exhibit a long-range FM order below the $T_C$. Adjusting these interactions is shown to preserve the large refrigerant capacity ($RC$) while tuning the $T_C$ over a wide temperature range, which is desirable for active magnetic refrigeration.

[*]Corresponding authors: faxiangqin@zju.edu.cn (F.X.Q.) phanm@usf.edu (M.H.P.)

## 1. Introduction

Refrigeration and air conditioning account for the dramatic increase in energy consumption in our modern society. Current refrigerators operate based on the gas compression/expansion (GCE) technique. However, the GCE technique has already reached its maximal cooling efficiency and imposed an environmental concern associated with the leakage of the harmful gases used as refrigerants (e.g., hydrofluorocarbons). Therefore, there is a pressing need for developing new, alternative cooling technologies, especially those utilizing solid-state materials [1,2]. Among the recently developed solid-state cooling technologies, magnetic refrigeration (MR) based on the magnetocaloric effect (MCE) has emerged as a promising alternative to the GCE refrigeration [2-5]. The MCE is a common phenomenon in magnetic materials, and the essence of which is a thermal effect caused by the change of magnetic moments order under an adiabatic condition [6,7]. The majority of research in this field is to design and engineer a cost-effective magnetic refrigerant that has a large adiabatic temperature change ($\Delta T_{ad}$) or a large magnetic entropy change ($\Delta S_M$) over a wide temperature range, resulting in the large refrigerant capacity (*RC*) [7-12]. A wide range of first-order magnetic transition (FOMT) materials have been reported to exhibit large or giant values of $\Delta S_M$, often accompanied by the structural phase transitions [13-16], but restricted to narrow temperature ranges, resulting in the moderate *RC* values, which, coupled with large hysteresis loss and poor mechanical properties, hinder their application in active magnetic refrigeration (AMR) [2,8,17]. In contrast, materials with a second-order magnetic transition (SOMT) possess moderate $\Delta S_M$ values but span over wide temperature ranges, resulting in the large RC values, which, together with negligible hysteresis loss, are desirable for AMR [18,19]. A typical example of such SOMT materials is Gadolinium (Gd) that possesses the desirable magnetocaloric properties, and was first used as a magnetic refrigerant in a laboratory scale cooling device for sub-room temperature AMR [20]. Gd is still the "most energy-efficient" magnetic refrigerant to date [21], and has been the benchmark to judge the cooling performance of newly discovered magnetocaloric materials [4,9,22,23].

For a practical cooling application perspective, however, the use of pure Gd in magnetic refrigerators is not economic and restricted to operate within the sub-room temperature regime [2]. Various efforts have been therefore made to overcome this limitation by alloying Gd with other materials, such as $Gd_5(Si_xGe_{1-x})_4$ compounds [9]. Recent studies have shown that a combination of alloying and amorphization not only reduces the manufacture cost of Gd and improves its mechanical strength, but also allows for the magnetocaloric tunability of the material over a wide range of temperatures [10,24-29]. The melt-extracted amorphous Gd alloy microwires have recently gained growing attention, owing to their large $\Delta S_M$ and $RC$ values achieved around the liquid nitrogen temperature desirable for use in nitrogen liquefaction [30-33]. So far, most research has focused on Gd-*TM*-Al based alloys [31,32,34-36] where *TM* refers to transition metal (mainly Fe, Co, Ni) and the adjustability of the Gd/*TM* ratio allows tuning the Curie temperature ($T_C$) of the alloy in a wide temperature range (70 K - 230 K). However, the underlying origin of the broad magnetocaloric behavior in these systems has remained elusive, due to the complexity of competing magnetic interactions that are present in the alloys. In addition to the Ruderman-Kittel-Kasuya-Yosida (RKKY)-typed ferromagnetic (Gd-Gd) interaction, other antiferromagnetic (Gd-*TM*) interactions coexist, contributing considerably to the overall magnetism of the Gd-*TM*-Al system. This raises an important question: *Can the ferromagnetic (FM) and antiferromagnetic (AFM) interactions be manipulated to drive the magnetocaloric tunability in Gd-TM-Al?*

To tackle this question, we have selected $Gd_{55}Co_{30}Ni_xAl_{15-x}$ ($x$ = 10, 5 and 0) amorphous microwires as a model system and investigated their magnetic and magnetocaloric properties. Based on our previous studies [10,23-25,32,33,37,38], we fixed the Gd/Co ratio in $Gd_{55}Co_{30}Ni_xAl_{15-x}$ while varying the Ni/Al ratio to probe effects of the Gd-Gd and Gd-Ni interactions on the magnetocaloric behavior. We demonstrate that the dilution effect of Al element on the ferromagnetic Gd-Gd interaction is more influential on reducing the $T_C$, while the change in saturation magnetization ($M_S$) is caused by the antiferromagnetic Gd-Ni interaction effect. A detailed analysis of critical exponents reveals the occurrence of a short-range ferromagnetic order below the $T_C$ for all samples investigated. It is worth mentioning that adjusting the FM and AFM interactions through the Al/Ni ratio can preserve the large *RC* value of the $Gd_{55}Co_{30}Ni_xAl_{15-x}$ microwires over a wide temperature range. This feature is very desirable for the design of a novel composite showing a "table-like" MCE performance for AMR [30,39].

**2. Experiment**

The amorphous alloy microwires with nominal compositions of $Gd_{55}Co_{30}Ni_xAl_{15-x}$ ($x$ = 10, 5 and 0) were fabricated by the melt-extraction method. The ingots were prepared from raw materials Gd (99.9%), Co (99.99%), Ni (99.99%) and Al (99.9%) in argon atmosphere by arc melting. The melt extraction process was performed using a copper wheel with diameter of 160mm and a 60° knife edge, with a linear velocity of the wheel rim fixed at30m/min and a feed rate of the molten material

at 90μm/s[24,40].

The microstructure was observed by field-emission scanning electron microscopy (SEM, SU-70) and the distribution of elements was determined by energy disperse spectroscopy (EDS). X-ray photoelectron spectroscopy (AXIS Supra, XPS) was used to detect the valence electron state of elements. Thermal analysis was performed in a differential scanning calorimeter (DSC) at the heating rate of 20 K/min from 323 K to 1273 K. The X-ray diffraction (XRD) patterns were obtained from MAXimaXRD-7000 using Cu-Ka radiation. The magnetic and magnetocaloric properties were characterized utilizing a commercial Physical Property Measurement System (PPMS-9T) from Quantum Design in a temperature range of 10 - 300 K and with magnetic field up to 5 T. The temperature interval is 5 K away from the $T_C$ and 3 K close to the $T_C$.

## 3. Results and discussion

### *3.1. Morphological and structural characterization*

The SEM image (Fig. 1(a)) shows the morphology of the $Gd_{55}Co_{30}Ni_{10}Al_5$ microwires. The SEM images of the other two samples are not shown here, due to the similarity. The mean diameter of the wires is around 50μm. The surface of the microwires is flat and smooth without grooves and fragments. Fig. 1(b) shows that the microwires possess a good roundness and are brittle. The surface-sweep EDS patterns of different atoms (Fig. 1(c)-(f)) show that the elements are evenly distributed across the cross section in the sample. The actual atomic percentage is determined to be 53.64% for Gd, 29.55% for Co, 9.51% for Ni, 7.30% for Al, which is nearly identical with the nominal composition. The actual atomic percentage of all the three samples is summarized in Tab. 1.

The XPS data reveal the valence electron states of Gd and Al, as shown in Fig. 2(a) and (b). The peak fitting is performed by XPSPEAK program. It should be stated that XPS reflects the binding energy of the elements on the surface of the material rather than the inside of the material, but it still has a reference significance [41], especially when the microwires were grinded into powder. The XPS data of Gd atom show a binding energy shift of 4d levels ($4d_{3/2}$ and $4d_{5/2}$), which is smaller than 0.5ev. This is likely related to the fact that the core levels of Gd are not sensitive to the chemical environment, when the content is constant and the transfer of charge from Gd to the *TM* d bands is smaller than one electron per Gd atom [42]. Different from the Gd atoms, the binding energy of the 2p level in Al atoms increases from 72.7eV to 73.2eV with Al addition. This indicates that the Al element obtains a larger coordination number as the Al content increases. Combined with the previous EDS data, the XPS data shows that Al was successfully incorporated into the amorphous alloy system.

The structural characterization of the microwire samples was performed by XRD, as shown in Fig. 3(a). The XRD pattern of the $Gd_{55}Co_{30}Ni_{15}$ microwires (without Al) shows a coexistence of amorphous and nanocrystalline phases [25,33,37]. The incorporation of Al in $Gd_{55}Co_{30}Ni_{15}$ has yielded a more adequate amorphous structure. It can be inferred that the increasing Al content can increase the glass forming ability (GFA) in this alloy. As a pseudo-binary alloy, the GFA of $Gd_{55}Co_{30}Ni_{15}$ is small. Increasing Al content can not only increase the number of components, but also increase the radius difference between atoms, resulting in a better GFA [43]. The DSC curves taken for the present microwire samples (Fig. 3b) prove the above point. With the Al percentage increasing from 5 to 15%, the glass transition temperature ($T_g$) and crystallization onset temperature ($T_x$) increase from 563 K to 566 K and from 583 K to 600 K, respectively. The corresponding supercooled liquid region ($\Delta T_x = T_x$-$T_g$) increases from 20 to 34 K. This indicates that the stability of the amorphous phase is greatly improved. This explains the binding energy increase of Al that strengthens the connection between atoms in the short-range ordered structure of the amorphous phase.

*3.2. Magnetic properties*

Fig. 4(a-c) shows the temperature dependence of magnetization *M(T)* for the microwire samples measured under an applied magnetic field of $\mu_0 H$ = 0.02 T. As can be seen in this figure, the *M*(*T*) curves present a broaden ferromagnetic to paramagnetic (FM-PM) phase transition around the Curie temperature. Values of $T_C$ ~ 127 K, 140 K, and 158 K for $Gd_{55}Co_{30}Al_{15}$, $Gd_{55}Co_{30}Ni_5Al_{10}$, and $Gd_{55}Co_{30}Ni_{10}A_5$, respectively, are determined by the minimum points of the derivative d*M*/d*T* vs. *T* curves (see the insets of Fig. 4). It can be seen that $T_C$ increases with increasing Ni content and decreasing Al content, and the *M*(*T*) curve becomes more and more broadened. Recently, our group has reported on the magnetic and magnetocaloric properties of GdCoAl alloy microwires [30,31,34]. It can be noticed that the obtained values of $T_C$ are quite low for these systems, such as $T_C$ ~ 101 K for $Gd_{60}Co_{15}Al_{25}$ [34], ~86 K for $Gd_{50}Co_{20}Al_{30}$, ~100 K for $Gd_{55}Co_{20}Al_{25}$, and ~109 K for $Gd_{60}Co_{20}Al_{20}$ [30]. In another study, the low value of $T_C$ (~ 80 K) was also reported for bulk metallic glass (BMG) $Gd_{55}Ni_{25}Al_{20}$ [26]. However, for active nitrogen liquefaction application, the $T_C$ of a magnetic refrigerant should range between 140 and 200 K[32]. In the present work, there is a higher percentage of the combination of two 3d transition metal elements (Ni/Co more than 30%), and the $T_C$ increases significantly, ~127 K for $Gd_{55}Co_{30}Al_{15}$, ~140 K for $Gd_{55}Co_{30}Ni_5Al_{10}$, and ~158

K for $Gd_{55}Co_{30}Ni_{10}Al_5$. In this alloy system, it appears that the Al addition plays a more dominant role than the 3d transition element in reducing the $T_C$.

The inverse of the magnetic susceptibility *vs* temperature curves, $\chi^{-1} = \mu_0 H/M$, specified from the $M(T)$ curves, are also shown in Fig. 4d. Because of the broad FM-PM phase transition, the $\chi^{-1}$-T curves are only linear at high temperature region. The Curie-Weiss temperatures $\theta$ and the effective magnetic moment $\mu_{\rm eff}$ values can be calculated by the Curie-Weiss law in the paramagnetic region, $\chi = \frac{C}{T-\theta}$, with the Curie constant defined by $C = \frac{N_A \mu_B^2}{3k_B}\mu_{eff}^2$, where $N_A$ = 6.022x10$^{23}$ mol$^{-1}$ is Avogadro's number, $\mu_B$ = 9.274x10$^{-21}$ emu is the Bohr magneton, and $k_B$ = 1.38016x10$^{-16}$ erg/K (in the CGS system of units) is Boltzmann constant. Fitting the linear region of the $\chi^{-1}$-T curves yields $\theta_1$ = 140.03 K, $C_1$ = 5.36 emu K mol$^{-1}$; $\theta_2$ = 151.37 K, $C_2$ = 4.98 emu K mol$^{-1}$; and $\theta_3$ = 169.84 K, $C_3$ = 4.33 emu K mol$^{-1}$ for $Gd_{55}Co_{30}Al_{15}$, $Gd_{55}Co_{30}Ni_5Al_{10}$, and $Gd_{55}Co_{30}Ni_{10}Al_5$, respectively. All Curie-Weiss temperatures are positive, which confirms the FM-PM phase transition. The difference between $\theta$ and $T_C$ is caused by the broad FM-PM phase transition, which indicates a structural disorder in the amorphous alloys [44]. The effective magnetic moments $\mu_{\rm eff}$ of all samples can be calculated via equation between $C$ and $\mu_{\rm eff}$: $\mu_{eff} = \left(\frac{3k_B C}{N_A}\right)^{1/2} = \sqrt{8C}\mu_B$. From this relationship, we obtained $\mu_{\rm eff}$ = 6.55$\mu_B$, 6.31$\mu_B$, and 5.89$\mu_B$ for the presently fabricated $Gd_{55}Co_{30}Al_{15}$, $Gd_{55}Co_{30}Ni_5Al_{10}$, and $Gd_{55}Co_{30}Ni_{10}Al_5$ microwires, respectively. These effective magnetic moments are smaller than the theoretically calculated moment of pure Gd (~7.94$\mu_B$), and the $\mu_{\rm eff}$ of $Gd_{55}Co_{30}Al_{15}$ (~6.55$\mu_B$) is also smaller than that of the previously reported $Gd_{60}Co_{15}Al_{25}$ microwires (~ 6.9$\mu_B$) [34]. The reduced value of $\mu_{\rm eff}$ is associated with the presence of AFM interactions between electrons of the 3d transition metal elements and the 4f rare-earth elements in the sample (Co/Ni vs. Gd). This AFM interaction is even more significant between Gd and Ni moments as compared to the AFM interaction between Gd and Co moments [26,45]. The enhancement of the AFM Gd-Ni interaction with Ni addition is found to reduce the saturation magnetization ($M_S$) in $Gd_{55}Co_{30}Ni_xAl_{15-x}$ microwires (Fig. S1).

Fig. 5(a,b,c) presents the temperature dependence of magnetization at different applied magnetic fields, $M(T)$ curves, from $\mu_0 H$ ~ 0.1 to 5T. As expected, all the $M(T)$ transitions of the samples become broadened with increasing applied magnetic field up to 5 T, which is typical behavior for SOMT materials. With increasing Ni and decreasing Al contents, the broadening of the $M(T)$ curves becomes more significant, due to the coexistence of competing FM and AFM interactions in these samples. Fig. 5(a-c) shows the filled 2D contour plots of the temperature and field dependence of the magnetization for all samples investigated. As can be seen clearly in these figures, a broadened phase transition from the magnetically disordered PM state to the ordered FM state is increasingly enlarged shown by the green area in Fig. 5d. The FM-PM phase transition starts to occur strongly at nearly the $T_C$ for all the microwire samples, suggesting the largest change in magnetic entropy ($\Delta S_M^{max}$) around this temperature region. The FM-PM phase transition extends over a wide range temperature (~50 K, 40 K, and 30 K for $Gd_{55}Co_{30}Ni_{10}Al_5$, $Gd_{55}Co_{30}Ni_5Al_{10}$, and $Gd_{55}Co_{30}Al_{15}$ microwires, respectively).

To evaluate the nature of the FM-PM phase transition, we have measured a set of isothermal magnetization curves of all the samples at different temperatures from 10 to 230 K with a temperature step interval between subsequent isotherms of $\delta T_1$=3 K for the temperature range near the $T_C$ and $\delta T_2$ = 5K for other temperature ranges with magnetic field changing from 0 to 5 T, as displayed in Fig. 6. The sweeping rate of the applied magnetic field is slow enough to ensure that the $M(\mu_0H)$ curves can be obtained in an isothermal process. This set of isothermal $M(\mu_0H)$ curves is an important measurement from which the magnetic and magnetocaloric properties of materials can be assessed. In Fig. 6, $M$ for all the samples show a large change in magnitude around the $T_C$. When temperature is decreased below the $T_C$, $M$ increases sharply and tends to saturate at a low field, indicating a good magnetic softness and low magnetic loss desirable for AMR.

*3.3. Magnetocaloric effect*

From the isothermal $M(\mu_0H)$ curves of all the microwire samples, the magnetic entropy change $\Delta S_M$ can be calculated using the thermal-dynamic Maxwell relation $\Delta S_M(T, \mu_0 H) = S_M(T, \mu_0 H) - S_M(T, 0) = \mu_0 \int_0^{H_{max}} \left(\frac{\partial M}{\partial T}\right)_H dH$, where $S_M(T, \mu_0 H)$ and $S_M(T, 0)$ are the magnetic entropy in a magnetic field $\mu_0 H$ and in the absence of magnetic field at constant temperature, respectively, $M$ is the magnetization, $\mu_0H$ is the applied magnetic field, and $T$ is the temperature, by integrating over the magnetic field. The temperature dependence of $-\Delta S_M$ at different applied magnetic fields ranging from $\mu_0\Delta H = 0 - 5$ T for these microwires are shown in Fig. 6. It can be clearly seen that the magnitude of $-\Delta S_M$ increased with an increase in the applied magnetic field and all the peaks of $-\Delta S_M$ appear to occur a maximum value near the $T_C$ as expected. For $\mu_o\Delta H = 5$ T, the $-\Delta S_M^{max}$ values are found to be ~ 9.71 J kg$^{-1}$ K$^{-1}$ at 127 K, ~ 8.91 J kg$^{-1}$ K$^{-1}$ at 140 K, and ~ 8.01 J kg$^{-1}$ K$^{-1}$ at 158 K for $Gd_{55}Co_{30}Al_{15}$, $Gd_{55}Co_{30}Ni_5Al_{10}$, and $Gd_{55}Co_{30}Ni_{10}Al_5$, respectively. These values of $-\Delta S_M^{max}$ are equivalent to those reported previously for GdCoAl microwires [30,31,34]. Adjusting the Al/Ni ratio in GdCoNiAl is found to increase the $T_C$ significantly, yet $-\Delta S_M^{max}$ simultaneously decreases. It is generally accepted that a magnetic cooling cycle can be estimated from $T_{hot}$ to $T_{cold}$ of the $-\Delta S_M$ ($T$) curve. Therefore, the $T_C$ of a magnetic refrigerant suitable for nitrogen liquefaction should be in the range of 140-200 K[32]. In this work, the $T_C$ values for $Gd_{55}Co_{30}Ni_5Al_{10}$ and $Gd_{55}Co_{30}Ni_{10}Al_5$ are in this desirable range ($T_C$ ~140 K for $Gd_{55}Co_{30}Ni_5Al_{10}$ and 158 K for $Gd_{55}Co_{30}Ni_{10}Al_5$). Therefore, it is advantageous to adjust the Al/Ni ratio in GdCoNiAl for significant increase of the $T_C$ while preserving the large value of $-\Delta S_M^{max}$. Additionally, the full width at half maximum (FWHM) of $-\Delta S_M$ ($T$,$\mu_0 H$) increases significantly upon the increasing of $\mu_o\Delta H$ ($\delta T_{FWHM}$ ~ 80 K, 85 K, and 90 K for $Gd_{55}Co_{30}Al_{15}$, $Gd_{55}Co_{30}Ni_5Al_{10}$, and $Gd_{55}Co_{30}Ni_{10}Al_5$ at 5T, respectively). The combination of such large $-\Delta S_M^{max}$ and large $\delta T_{FWHM}$ results in the large refrigerant capacity (RC) or the large relative cooling power (RCP), an important figure-of-merit for assessing the cooling efficiency of a magnetic refrigerant.

The RC values of all the samples are calculated by integrating the area under the $-\Delta S_M(T)$ curves using the temperatures at half maximum of the peaks [4]: $\mathrm{RC} = \int_{\mathrm{T_{hot}}}^{\mathrm{T_{cold}}} -\Delta S_\mathrm{M}(T)\mathrm{d}T$, where $T_{cold}$ and $T_{hot}$ are the onset and offset temperatures of $\delta T_{FWHM}$, respectively. The RCP shows an amount of heat transfer between the hot and cold sides in an ideal refrigeration cycle. The RCP values of the present samples can be evaluated using Wood and Potter's method: $RCP = -\Delta S_M^{max} \times \delta T_{FWHM}$ [9]. The RC($\mu_0 H$) and RCP($\mu_0 H$) values for the microwire samples are described in Fig. S2 for different field changes, $\mu_o\Delta H$. As can be seen in this figure, both the RC and RCP values increase significantly with increasing $\mu_o\Delta H$. The RC values are approximately 3/4 of the RCP values, which is reasonable for a typical broad peak in $-\Delta S_M(T)$ for a magnetic amorphous material. For $\mu_o\Delta H$ = 5 T, the RC and RCP values are calculated to be ~ 572 J $kg^{-1}$ and 702.5 J $kg^{-1}$; 531.8 J $kg^{-1}$ and 668 J $kg^{-1}$; 534.6 J $kg^{-1}$ and 681 J $kg^{-1}$ for $Gd_{55}Co_{30}Al_{15}$, $Gd_{55}Co_{30}Ni_5Al_{10}$, and $Gd_{55}Co$- $_{30}Ni_{10}Al_5$, respectively. An interesting feature of note is that the RC and RCP values are almost equal for all the compositions, indicating a balance between the decrease of $T_C$ and the increase of $-\Delta S_M(T)$ in $Gd_{55}Co_{30}Ni_xAl_{15-x}$ through adjusting the Al/Ni ratio, beneficial for the design of a novel "table-like" MCE composite for AMR.

Based on the above findings, the roles Al plays in the Gd-*TM*-Al systems are (i) increasing the GFA and thermal stability of the amorphous host and (ii) decreasing the $T_C$ of the material. The former can be attributed to the increase of the mixing entropy in an alloy system [43]. As Al is a nonmagnetic element, the magnetic moment is zero. The effect of adding this nonmagnetic element to a magnetic system is *dilution*, which, in turn, increases the Gd-Gd distance, thus weakening the FM Gd-Gd interaction leading to the overall decrease of $T_C$. On the other hand, the enhancement of the $M_S$ is mainly originated from the decrease of the Ni content that weakens the AFM Gd-Ni interaction.

*3.4. Critical exponents*

In a SOMT material, the critical behavior of the FM-PM phase transition can be specified by a set of three critical exponents, $\beta$, $\gamma$, and $\delta$ [46], which have been correlated with the magnetocaloric behavior of SOMT materials [7,47-50]. These critical exponents, $\beta$, $\gamma$, and $\delta$, are associated with the spontaneous magnetization $M_S(T)$, the thermal dependence of the initial inverse susceptibility $\chi_0^{-1}(T)$, and the field dependence of the magnetization of the critical isotherm, respectively [46]. These three exponents can be defined by using the following power-law relations:

$$M_S(\mathrm{T}) = M_0(-\varepsilon)^{\beta} \qquad \varepsilon < 0 \qquad T < T_C \tag{1}$$

$$\chi_0^{-1}(\mathrm{T}) = \left(\frac{H_0}{M_0}\right)\varepsilon^{\gamma} \qquad \varepsilon > 0 \qquad T > T_C \tag{2}$$

$$H = DM^{\delta} \qquad \varepsilon = 0 \qquad T = T_C \qquad (3)$$

where $\varepsilon$ is the reduced temperature, $\varepsilon = (T - T_C)/T_C$, and $M_0$, $H_0$, $D$ are the critical amplitudes.

Depending on the Arrott-Noakes equation of state $\left(\frac{H}{M}\right)^{\frac{1}{\gamma}} = A\varepsilon + BM^{\frac{1}{\beta}}$ [51](4), where A and B are material dependent parameters, the Curie temperature $T_C$ for all amorphous alloy microwires can be redefined by using modified Arrott plot (MAP) method. To use this MAP approach, the isothermal magnetization $M(\mu_o H)$ data is reformulated, using initial values of $\beta$ and $\gamma$, into $M^{1/\beta}$ $[(\mu_0 H/M)^{1/\gamma}]$ according to the equation (4). The correct critical exponents are those that linearize $M^{1/\beta}$ vs. $(\mu_0 H/M)^{1/\gamma}$, and $T_C$ is specified from the critical isotherm which passed through the origin. The initial $\beta$ and $\gamma$ values can be determined based on one of four theoretical models: the mean-field model ($\beta$ = 0.5, $\gamma$ = 1.0) , the 3D Ising model ($\beta$ = 0.325, $\gamma$ = 1.240) , the 3D Heisenberg model ($\beta$ = 0.365, $\gamma$ = 1.336) , and the tricritical mean field model ($\beta$ = 0.25, $\gamma$ = 1) [52,53] so as to establish the corresponding MAP for studying the critical behavior near the FM-PM phase transition of all amorphous alloy microwires. The spontaneous magnetization $M_S(T)$ and the initial inverse susceptibility $\chi_0^{-1}(T)$ were specified from the intercepts of the linear extrapolation of $M^{1/\beta}$ and $(\mu_0 H/M)^{1/\gamma}$ at the high field isotherms of the modified-Arrott plots, respectively. The critical exponents $\beta$ and $\gamma$ are extracted by fitting $M_S(T)$ data with the relation $M_S \propto (-\varepsilon)^{\beta}$ in equation (1) and $\chi_0^{-1}$(T) data with the relation $\chi_0^{-1} \propto \varepsilon^{\gamma}$ in equation (2), respectively. Using the new $\beta$ and $\gamma$ values this process is repeated until the isotherm that passes through the origin gives $T = T_C$ and the critical exponent $\beta$ and $\gamma$ values reach the stable values.

Fig. 7(a-c) presents the final modified-Arrott plots, in which the isotherms achieve good linearity with $\beta_{MAP}$ = 0.462±0.001, 0.458±0.002, 0.490±0.001, $\gamma_{MAP}$ = 1.064±0.004, 1.065±0.005,

1.029±0.002, and $T_{C\text{-}MAP}$ = 132.02 K, 143.94 K, 162.51 K for $Gd_{55}Co_{30}Al_{15}$, $Gd_{55}Co_{30}Ni_5Al_{10}$, and $Gd_{55}Co_{30}Ni_{10}Al_5$, respectively. In accordance with the statistical theory, these critical exponents $\beta$, $\gamma$ and $\delta$ must persuade the Widom scaling relation: $\delta = 1 + \frac{\gamma}{\beta}$ [54] (5). From the Widom scaling equation, the critical exponent $\delta$ can be calculated, $\delta$= 3.303±0.013, 3.325±0.11, 3.104±0.004 for $Gd_{55}Co_{30}Al_{15}$, $Gd_{55}Co_{30}Ni_5Al_{10}$, and $Gd_{55}Co_{30}Ni_{10}Al_5$, respectively.

Moreover, this set of critical exponents $\beta$, $\gamma$ and $\delta$ can also be specified by the Kouvel – Fisher (KF) method by the way of reformulating equations (1) and (2) in a different direction [55]:

$$M_S(\mathrm{T})\left[\frac{dM_S(\mathrm{T})}{dT}\right]^{-1} = \frac{T - T_C}{\beta} \tag{6}$$

$$\chi_0^{-1}(\mathrm{T})\left[\frac{d\chi_0^{-1}(\mathrm{T})}{dT}\right]^{-1} = \frac{T - T_C}{\gamma} \tag{7}$$

By this way, $M_S(T)\left[\frac{\mathrm{d}M_S(T)}{\mathrm{d}T}\right]^{-1}$ vs. $T$ and $\chi_0^{-1}(T)\left[\frac{\mathrm{d}\chi_0^{-1}(T)}{\mathrm{d}T}\right]^{-1}$ vs. $T$ are plotted in the same figures for all microwires, which result in straight lines with slopes of $1/\beta$, $1/\gamma$, and intercept the temperature axis at $T_C$. The exponents obtained via the linear fit are used to construct a new Arrott-Noakes plot. The procedure is repeated until the critical values converge. Fig. 7 (d-f) displays the final values of $M_S(T)\left[\frac{\mathrm{d}M_S(T)}{\mathrm{d}T}\right]^{-1}(T)$ and $\chi_0^{-1}(T)\left[\frac{\mathrm{d}\chi_0^{-1}(T)}{\mathrm{d}T}\right]^{-1}(T)$ and their linear fitting curves for all samples. For temperatures below $T_C$, the result for critical exponent $\beta$ is $\beta_{KF}$ = 0.463 ± 0.007, 0.461 ± 0.007, 0.488 ±0.005 and the Curie temperature is $T_{C1}$ = 132.06 ± 0.28, 143.93 ± 0.27, 162.43 ± 0.21 for $Gd_{55}Co_{30}Al_{15}$, $Gd_{55}Co_{30}Ni_5Al_{10}$, and $Gd_{55}Co_{30}Ni_{10}Al_5$, respectively. For temperatures above $T_C$, the result for the critical exponent $\gamma$ is $\gamma_{KF}$ = 1.074 ± 0.009, 1.071 ± 0.010, 1.025 ± 0.004 and the Curie temperature is $T_{C2}$= 131.97 ± 0.14, 143.93 ± 0.15, 162.57 ± 0.04 for $Gd_{55}Co_{30}Al_{15}$, $Gd_{55}Co_{30}Ni_5Al_{10}$, and $Gd_{55}Co_{30}Ni_{10}Al_5$, respectively. The critical exponent $\delta$ can be defined by using Widom scaling equation (5), $\delta_{KF}$ = 3.320 ± 0.023, 3.323 ± 0.024, 3.100 ± 0.014 for $Gd_{55}Co_{30}Al_{15}$, $Gd_{55}Co_{30}Ni_5Al_{10}$, and $Gd_{55}Co_{30}Ni_{10}Al_5$, respectively. The critical exponents for all the microwire samples obtained using the MAP and KF methods are listed in Tab. 2 for a comparison purpose. A good agreement between these two methods confirms that the calculated values of the critical exponents are reliable.

According to the power-law relation (eq. 3) with $(\varepsilon = 0, T = T_C)$, the critical exponent $\delta$ can also be defined by taking two-sided natural logarithms of equation. From that point, log-log plots of applied magnetic field dependence of magnetization ln$M$-ln($\mu_0 H$) at temperatures in the vicinity of $T_C$ for all microwires samples are shown in Fig. 8. From equation (3), the critical exponent $\delta$ can be defined from the inverse slope of the critical isotherm analysis (CIA). ln$M$-ln($\mu_0 H$) linear fitting curves of the $T_C$ = 132 K, 144 K, 163 K isotherms yield $\delta_{CIA}$ = 3.347 ± 0.009, 3.334 ± 0.008, 3.072 ± 0.008 for $Gd_{55}Co_{30}Al_{15}$, $Gd_{55}Co_{30}Ni_5Al_{10}$, and $Gd_{55}Co_{30}Ni_{10}Al_5$, respectively. These $\delta_{CIA}$ values are close to those determined from the Widom relation (eq. 5) based on the results of the MAP method and the KF method, as summarized in Tab. 2. All the critical exponents have been deduced from the different methods are almost equal.

For SOMT materials, there is an additional critical exponent, $n$, can be specified by the magnetic field dependence of $\Delta S_M^{max}$, in which follows a scaling law [56]: $\Delta S_M^{max} \propto \mu_0 H^n$ (8), where $n$ is a scaling exponent for the field dependence of the peak in the magnetic entropy change. $n$ is extracted from fitting with equation (8), as $n$ = 0.727±0.0013; 0.718±0.0015; 0.695±0.0011 for $Gd_{55}Co_{30}Ni_{10}Al_5$, $Gd_{55}Co_{30}Ni_5Al_{10}$, and $Gd_{55}Co_{30}Al_{15}$, respectively. Rescaling the field axis to produce a plot of $\Delta S_M^{max}$ vs. $(\mu_0 H)^n$ with $n$ = 0.695, 0.718, 0.727 reveals the expected relationship (shown in Fig. 9). A relationship between the scaling exponent $n$, the magnetization exponent $\beta$, and susceptibility exponent $\gamma$ can be described as: $n = 1 + \frac{\beta - 1}{\beta + \gamma}$ [57]. From those critical exponents $\beta$ and $\gamma$ calculated from MAP and KF methods, $n_{MAP}$ and $n_{KF}$ are defined by using this relationship, yields $n_{MAP}$ = 0.647, 0.644, 0.664 and $n_{KF}$ = 0.651, 0.648, 0.662. The calculated critical exponents are also listed in Tab. 2 for the comparison purpose.

To reconfirm the validity of all the critical exponents determined from the MAP and KF methods, these exponents can be tested by the scaling analysis via the static-scaling hypothesis, which relates to the magnetization $M$ and applied magnetic field $\mu_0 H$. The isothermal magnetization data is rescaled based on the renormalized equation of state [58]:

$$m = f_{\pm}(h) \tag{10}$$

$$h/m = \pm a_{\pm} + b_{\pm} m^2 \tag{11}$$

where the plus and minus signs depict the temperature ranges above and below $T_C$, respectively; $m \equiv |\varepsilon|^{-\beta} M(H,\varepsilon)$ and $h \equiv |\varepsilon|^{-\beta\delta} H$ are the renormalized magnetization and magnetic field, respectively. The expression $f_{\pm}(h) = M(h, \varepsilon / |\varepsilon| = \pm 1)$ determines two universal curves onto which the rescaled magnetization data should collapse above and below $T_C$ [58]. Fig. S3 presents the good collapse of the rescaled magnetization data based on equation (10) and (11). These results confirm that all the calculated critical exponents are correct.

Tab. 2 summarizes all the critical exponents ($\beta$, $\gamma$, $\delta$, and $n$) for all the microwires calculated by various methods in this work, in comparison with those obtained from previous studies as well as theoretical methods. The exponents, $\beta$ (0.463, 0.461, 0.488), $\gamma$ (1.074, 1.071, 1.025), $\delta$ (3.347, 3.334, 3.072), and $n$ (0.651, 0.648, 0.662) are all close to those of the mean-field model ($\beta$ = 0.5, $\gamma$ = 1.0, $\delta$

= 3.0, and $n$ = 0.67). However, the values of $\beta$ (0.463; 0.461; 0.488) are smaller than that deduced from the mean-field model (between the mean-field model and the 3D- Heisenberg model). This once again suggests that the additional presence of the AFM (Gd-Ni, Gd-Co) interactions in the $Gd_{55}Co_{30}Ni_xAl_{15-x}$ microwires could hinder the occurrence of a long-range ferromagnetic order below the $T_C$ [37,48,59] (Fig. S4).

## 4. Conclusion

The magnetic and magnetocaloric properties of the $Gd_{55}Co_{30}Ni_xAl_{15-x}$ ($x$ = 10, 5 and 0) amorphous microwires have been systematically investigated. We show the coexistence of ferromagnetic (Gd-Gd) and antiferromagnetic (Gd-Co, Gd-Ni) interactions which all contribute to the magnetic and magnetocaloric response of the system. The dilution effect of Al element on the FM Gd-Gd interaction is responsible for reducing the Curie temperature ($T_C$), while the enhancement of saturation magnetization ($M_S$) is resulted from the weakened AFM Gd-Ni interaction due to the increased Ni content. The additional presence of the AFM (Gd-Co, Gd-Ni) interactions hinders the system to exhibit a long-range FM order below the $T_C$. Adjusting FM and AFM interactions to preserve the large *RC* while tuning the $T_C$ over a wide temperature range in a multiple magnetic interacting system is a feasible approach for the design of energy-efficient magnetic refrigerants for active magnetic refrigeration technology.

**Acknowledgments**

This work is supported by ZJNSF No. LR20E010001and Aeronautical Science Foundation 2019ZF076002 and Zhejiang Provincial Key Research and Development Program (2019C01121).

**Table 1.** The actual atomic percentage of the $Gd_{55}Co_{30}Ni_xAl_{15-x}$ ($x$ = 10, 5 and 0) microwire samples

| **Samples** | **Gd** | **Co** | **Ni** | **Al** |
|---|---|---|---|---|
| $Gd_{55}Co_{30}Ni_{10}Al_5$ | 53.64% | 29.55% | 9.51% | 7.30 |
| $Gd_{55}Co_{30}Ni_5Al_{10}$ | 53.38% | 29.80% | 4.86% | 11.96% |
| $Gd_{55}Co_{30}Al_{15}$ | 52.54% | 30.01% | 0 | 17.46% |

**Table 2.** Curie temperature $T_C$ and critical exponents of $Gd_{55}Co_{30}Ni_xAl_{15-x}$ ($x$ = 10, 5 and 0) in comparison those of the theoretical models and other materials.

| **Material/ Model** | **Method** | $T_C$**(K)** | $\beta$ | $\gamma$ | $\delta$ | $n$ | **Ref.** |
|---|---|---|---|---|---|---|---|
| Mean field | Theory | - | 0.5 | 1.0 | 3.0 | 0.67 | [52] |
| Tricritical Mean field | Theory | - | 0.25 | 1.0 | 5.0 | - | [53] |
| 3D Heisenberg | Theory | - | 0.365 | 1.336 | 4.80 | - | [52] |
| 3D Ising | Theory | - | 0.325 | 1.241 | 4.82 | - | [52] |
| $Gd_{55}Co_{30}Al_{15}$ microwires | MAP | 132.02 | 0.462±0.001 | 1.064±0.004 | 3.303±0.013 | 0.647 | |
| | KF (below $T_C$) | 132.06 ±0.28 | 0.463 ±0.007 | - | 3.320 ±0.023 | 0.651 | |
| | KF (above $T_C$) | 131.97 | - | 1.074 ±0.009 | | | |
| | CIA | - | - | - | 3.347 ±0.009 | - | |
| | Eq.(8) | - | - | - | - | 0.695 | |
| $Gd_{55}Co_{30}Ni_5Al_{10}$ microwires | MAP | 143.94 | 0.458±0.002 | 1.065±0.005 | 3.325±0.11 | 0.644 | |
| | KF | 143.93 | 0.461 | - | 3.323 | 0.648 | |
| | (below $T_C$) | ±0.27 | ±0.0.007 | | ±0.024 | | |
| | KF (above $T_C$) | 143.93 ±0.15 | - | 1.071 ±0.010 | | | |
| | CIA | - | - | - | 3.334 ±0.008 | - | |
| | Eq.(8) | - | - | - | - | 0.718 | |
| $Gd_{55}Co_{30}Ni_{10}Al_5$ microwires | MAP | 162.51 | 0.490±0.001 | 1.029±0.002 | 3.104±0.004 | 0.664 | |
| | KF (below $T_C$) | 162.43 ±0.21 | 0.488 ±0.005 | - | 3.100 ±0.014 | 0.662 | |
| | KF (above $T_C$) | 162.57 ±0.0.04 | - | 1.025 ±0.004 | | | |
| | CIA | - | - | - | 3.072 ±0.008 | - | |
| | Eq.(8) | - | - | - | - | 0.727 | |
| $Gd_{60}Fe_{20}Al_{20}$ microwires | MAP | 202.15 | 0.723 | 1.246 | 2.724 | 0.859 | [32] |
| $Gd_{60}Co_{15}Al_{25}$ | MAP | 105 | 0.432 | 1.244 | 3.51 | 0.738 | [34] |

**Figure Captions:**

**Fig. 1.** (a) SEM image of a $Gd_{55}Co_{30}Ni_{10}Al_5$ microwire; (b) cross-sectional SEM image of the microwire; (c)-(f) the EDS patterns for different elements of the microwire in a surface-sweep mode.

**Fig. 2.** XPS spectra of the $Gd_{55}Co_{30}Ni_xAl_{15-x}$ ($x$ = 10, 5 and 0) samples for (a) Gd and (b) Al atoms.

**Fig. 3.** XRD patterns (a) and DSC curves (b) for the $Gd_{55}Co_{30}Ni_xAl_{15-x}$ ($x$ = 10, 5 and 0) samples.

**Fig. 4.** (a)-(c) Temperature dependence of magnetization at a field of 0.2 T for the $Gd_{55}Co_{30}Ni_xAl_{15-x}$ ($x$ = 10, 5 and 0) samples; the insets show the $dM/dT$ versus $T$ curves, respectively; (d) Temperature dependence of the inverse magnetic susceptibility and the Curie-Weiss fit for the $Gd_{55}Co_{30}Ni_xAl_{15-x}$ ($x$ = 10, 5 and 0) samples.

**Fig. 5.** (a)-(c) Temperature dependence of magnetization at magnetic field from 0.1 to 5 T for $Gd_{55}Co_{30}Ni_xAl_{15-x}$ ($x$ = 10, 5 and 0) microwires; (d)-(f) Filled two-dimensional (2D) contour plots of temperature and applied magnetic field dependence of magnetization for $Gd_{55}Co_{30}Ni_xAl_{15-x}$ ($x$ = 10, 5 and 0) microwires.

**Fig. 6.** (a)-(c) Magnetic field dependence of magnetization from 10 K-230 K for $Gd_{55}Co_{30}Ni_xAl_{15-x}$ ($x$ = 10, 5 and 0) microwires; (d)-(f) Temperature dependence of magnetic entropy change ($-\Delta S_M$) for different applied field changes (0.1 to 5 T) for $Gd_{55}Co_{30}Ni_xAl_{15-x}$ ($x$ = 10, 5 and 0) microwires.

**Fig. 7.** (a)-(c) Arrott–Noakes plots of the $Gd_{55}Co_{30}Ni_xAl_{15-x}$ ($x$ = 10, 5 and 0) microwires; (d)-(f) KF plots for the $Gd_{55}Co_{30}Ni_xAl_{15-x}$ ($x$ = 10, 5 and 0) microwires.

**Fig. 8.** (a)-(c) ln(M) versus ln(H) for temperatures near $T_C$ for the $Gd_{55}Co_{30}Ni_xAl_{15-x}$ ($x$ = 10, 5 and 0) microwires.

**Fig. 9.** $-\Delta S_M^{max}$ versus $H$ profiles for the $Gd_{55}Co_{30}Ni_xAl_{15-x}$ ($x$ = 10, 5 and 0) samples

**Figure 1**

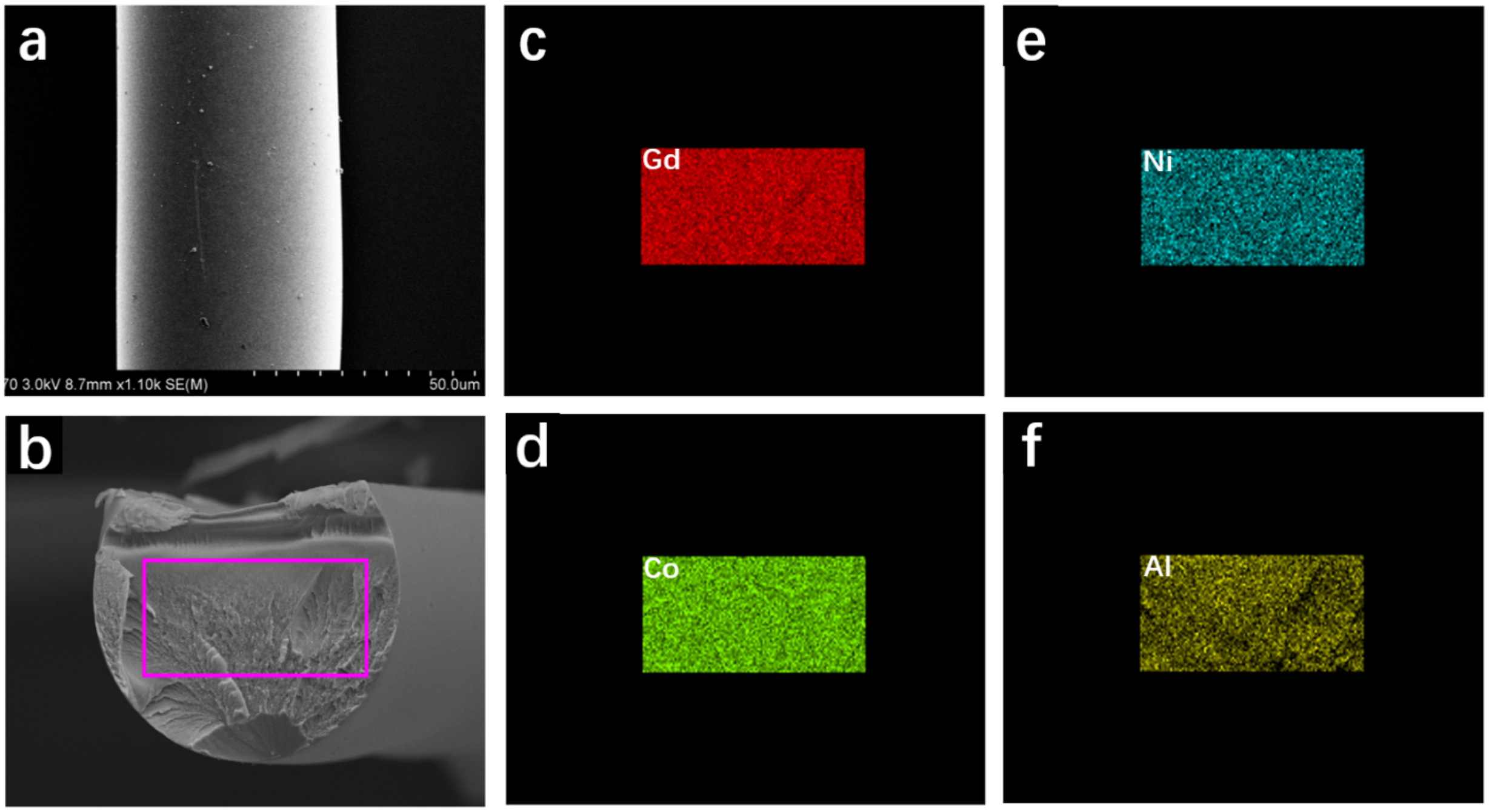

**Figure 2**

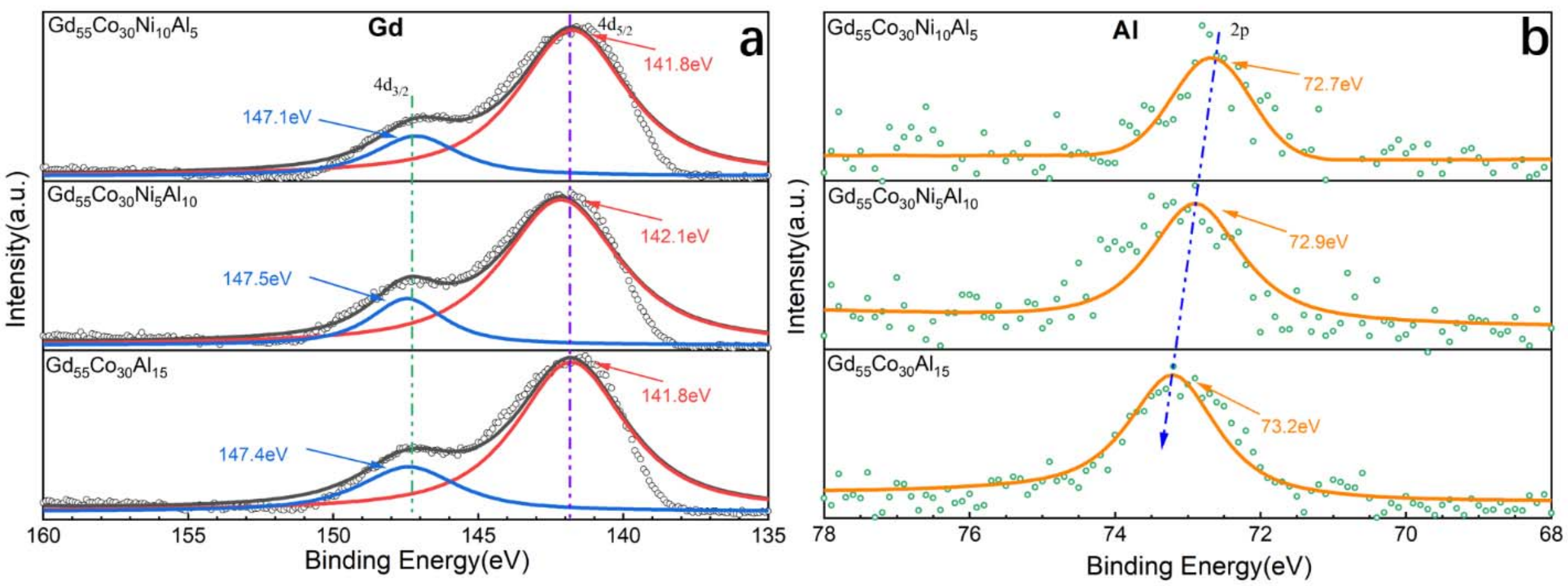

**Figure 3**

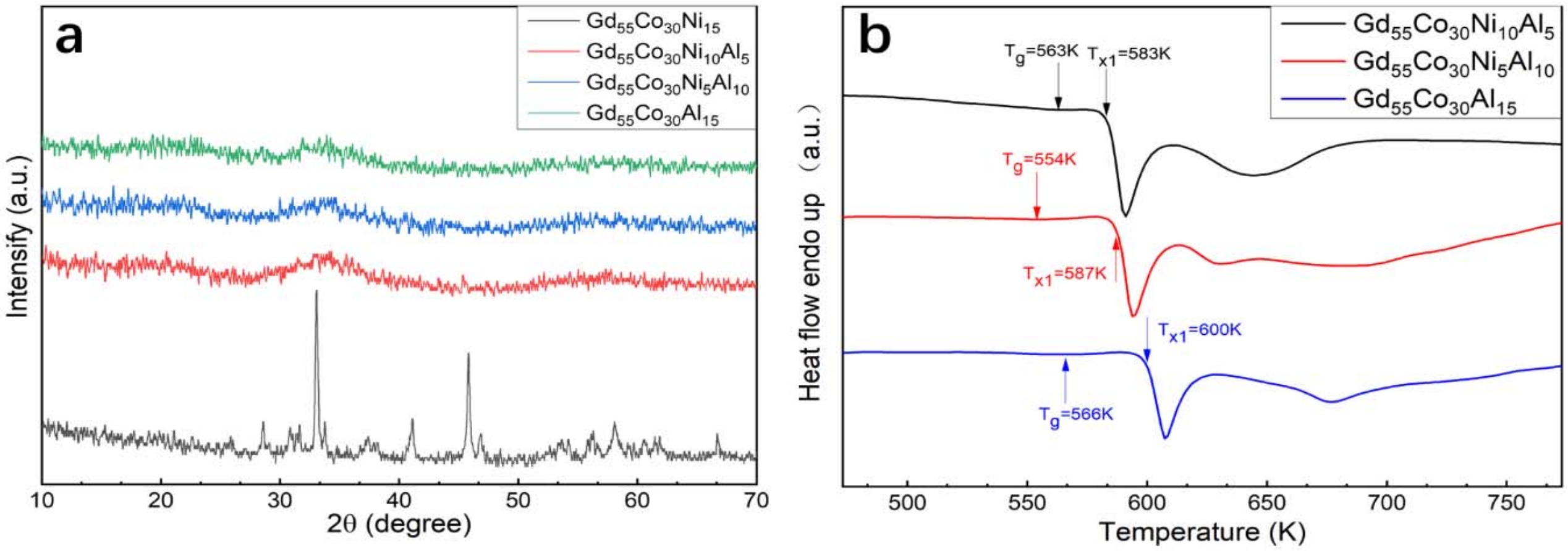

**Figure 4**

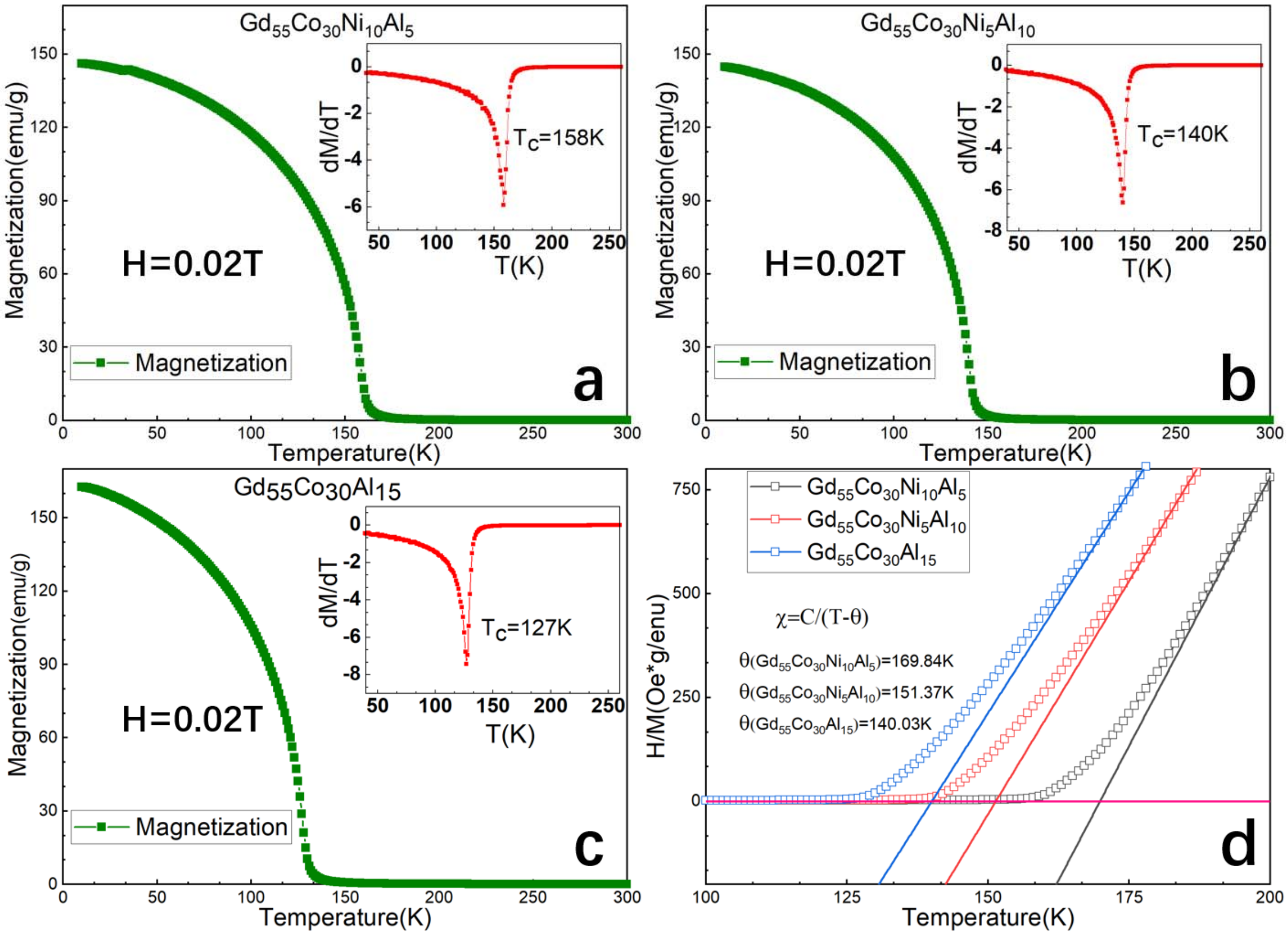

**Figure 5**

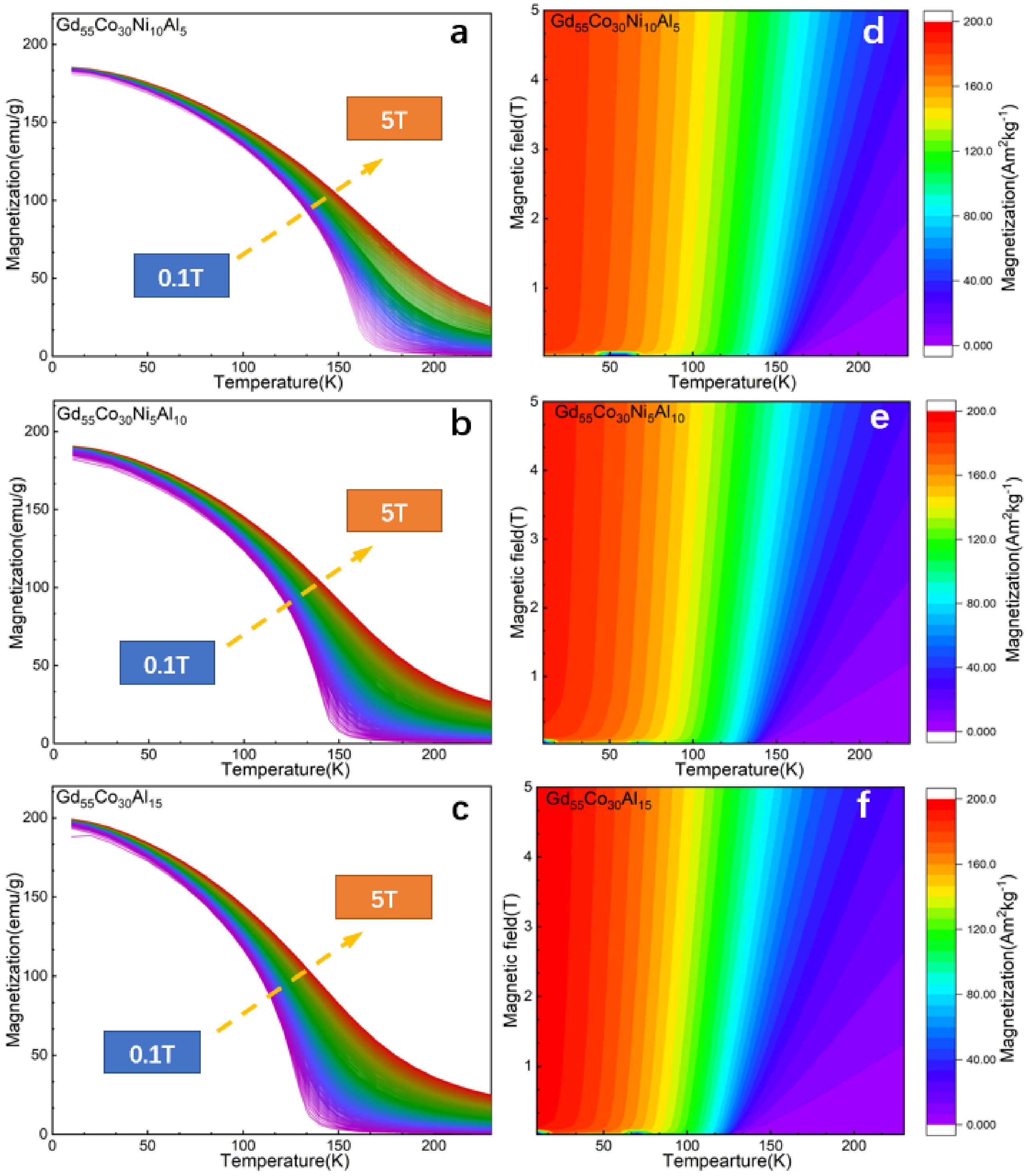

**Figure 6**

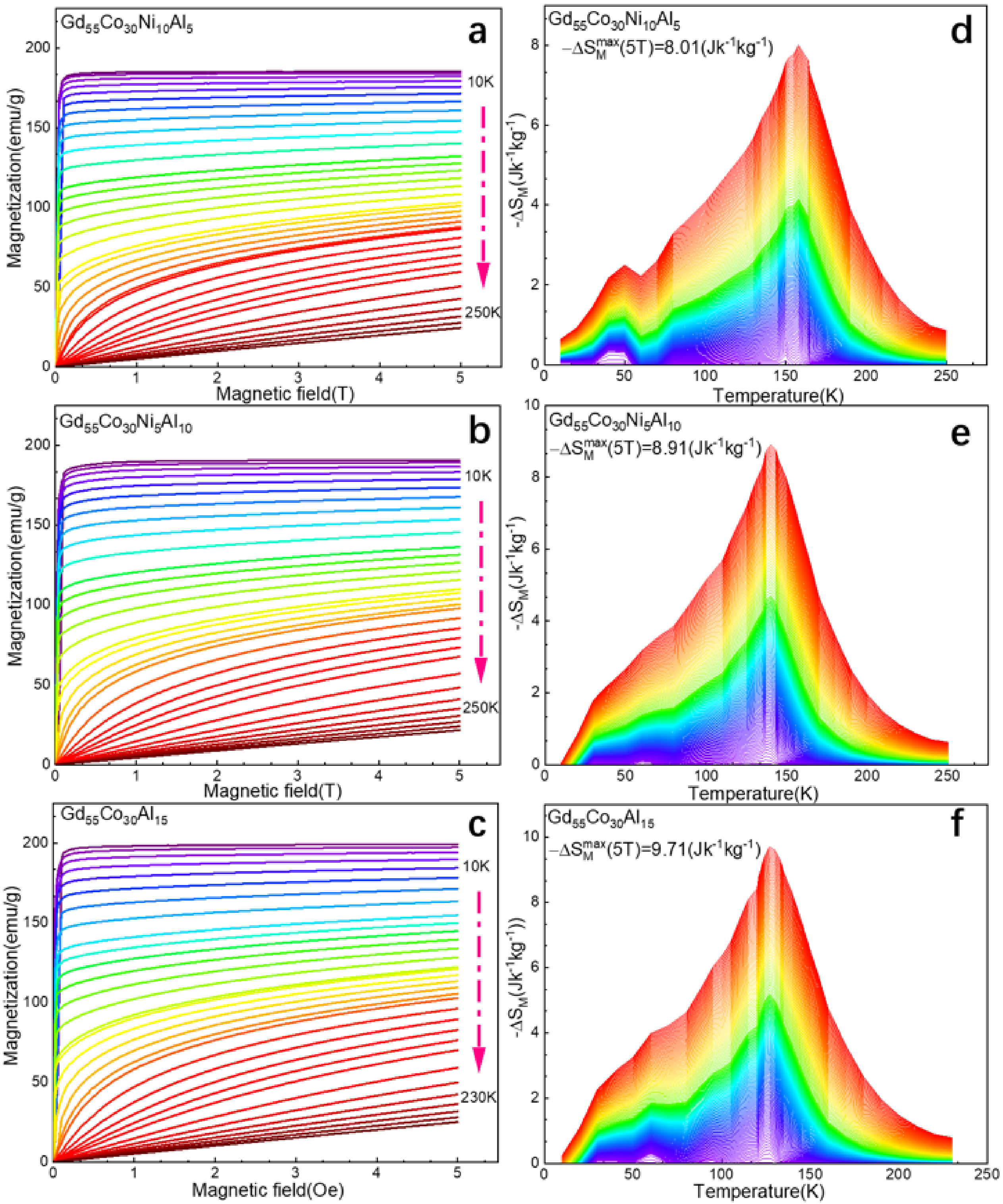

**Figure 7**

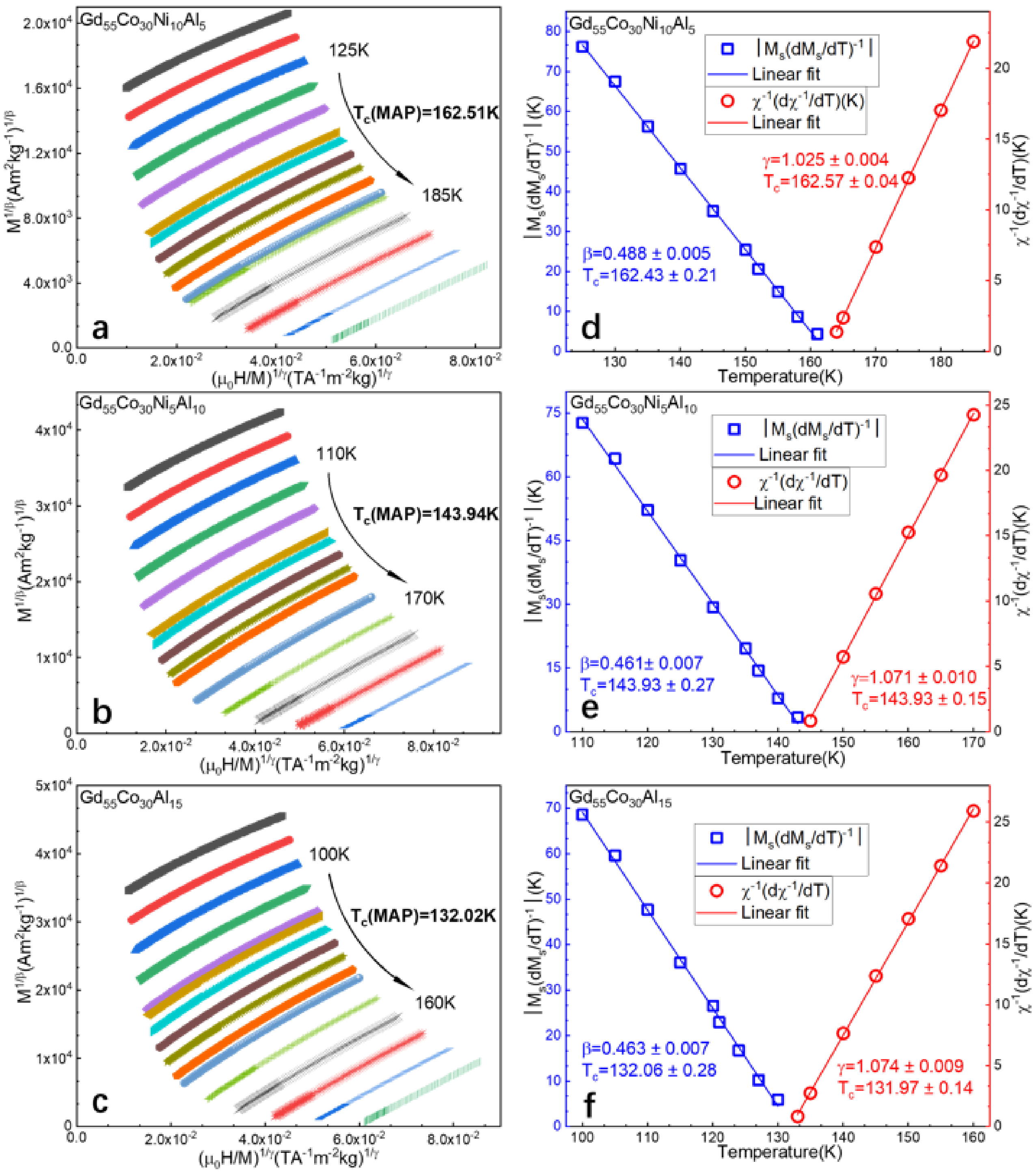

Gd55Co30Ni10Al5
125K
Tc(MAP)=162.51K
185K
a
M1/β(Am2kg-1)1/β
(μ0H/M)1/γ(TA-1m-2kg)1/γ
Gd55Co30Ni10Al5
|Ms(dMs/dT)-1|
Linear fit
χ-1(dχ-1/dT)(K)
Linear fit
γ=1.025 ± 0.004
Tc=162.57 ± 0.04
β=0.488 ± 0.005
Tc=162.43 ± 0.21
d
Temperature(K)
Gd55Co30Ni5Al10
110K
Tc(MAP)=143.94K
170K
b
Gd55Co30Ni5Al10
|Ms(dMs/dT)-1|
Linear fit
χ-1(dχ-1/dT)
Linear fit
β=0.461± 0.007
Tc=143.93 ± 0.27
γ=1.071 ± 0.010
Tc=143.93 ± 0.15
e
Gd55Co30Al15
100K
Tc(MAP)=132.02K
160K
c
Gd55Co30Al15
|Ms(dMs/dT)-1|
Linear fit
χ-1(dχ-1/dT)
Linear fit
β=0.463 ± 0.007
Tc=132.06 ± 0.28
γ=1.074 ± 0.009
Tc=131.97 ± 0.14
f

**Figure 8**

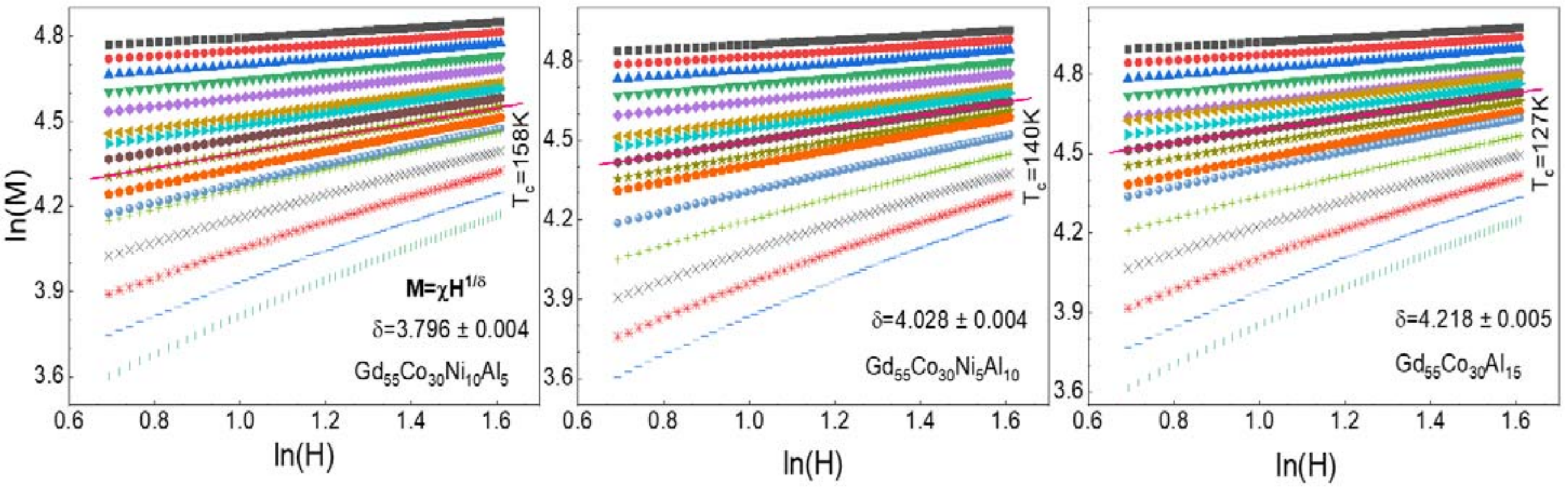

**Figure 9**

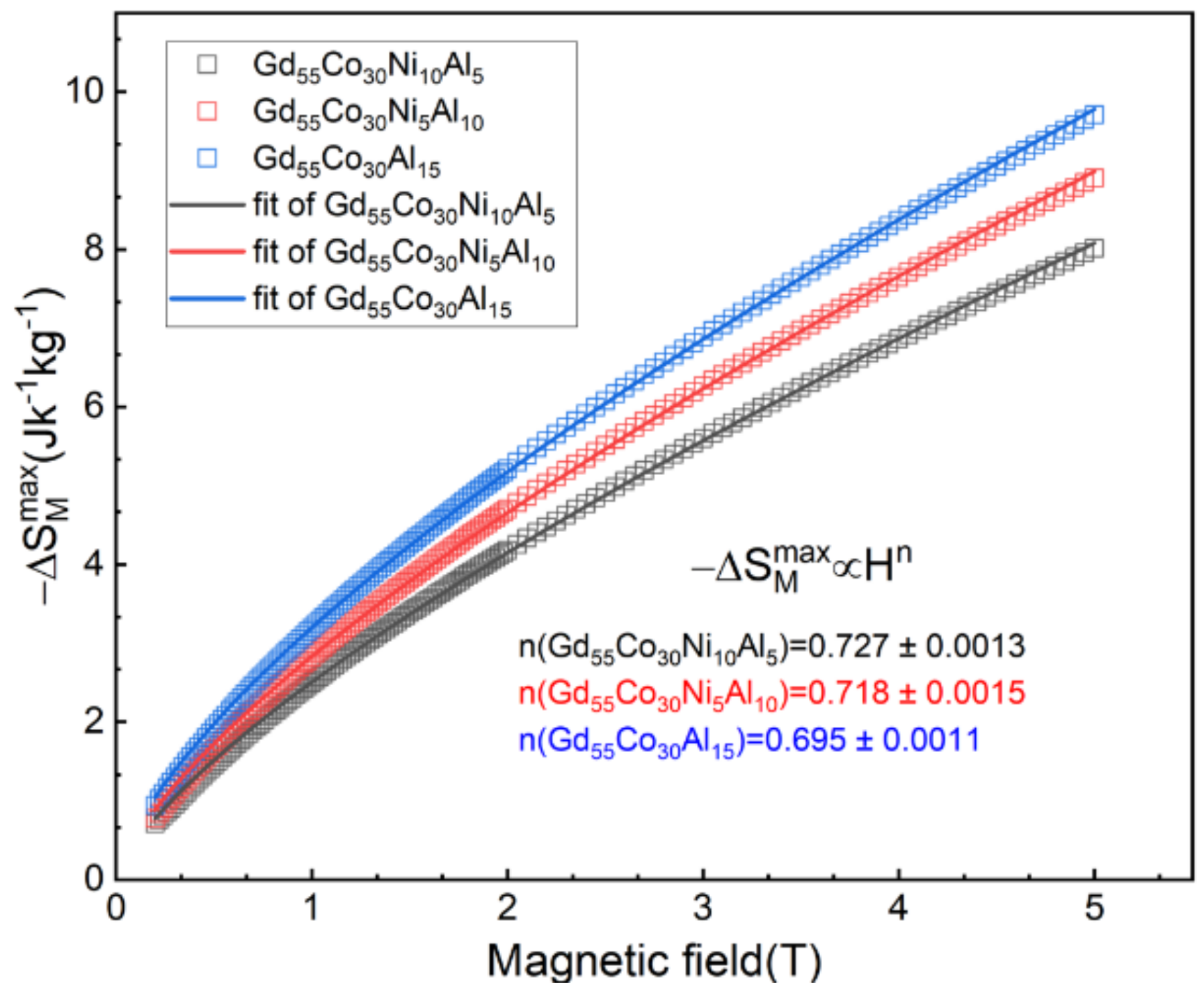